\documentclass[structabstract]{aa}
\usepackage{txfonts}
\usepackage{natbib}
\usepackage{graphicx}
\usepackage{color}
\definecolor{red}{rgb}{1,0,0}
\definecolor{blue}{rgb}{0,0,1}

\begin{document}

\title{The nearby eclipsing stellar system $\delta$ Velorum}
\subtitle{IV. Differential astrometry with VLT/NACO at the 100 microarcsecond level
\thanks{Based on observations made with ESO telescopes at the La Silla Paranal Observatory, under ESO programs 081.D-0109(A), 081.D-0109(B), 382.D-0065(A), and 282.D-5006(A).}}
\titlerunning{Differential astrometry of $\delta$ Velorum with VLT/NACO}
\authorrunning{P. Kervella et al.}
\author{
P. Kervella\inst{1}
\and
A. M\'erand\inst{2}
\and
M. G. Petr-Gotzens\inst{3}
\and
T. Pribulla\inst{4}
\and
F. Th\'evenin\inst{5}
}
\offprints{P. Kervella}
\mail{Pierre.Kervella@obspm.fr}
\institute{
LESIA, Observatoire de Paris, CNRS\,UMR\,8109, UPMC, Universit\'e Paris Diderot, 5 place Jules Janssen, 92195 Meudon, France
\and
European Southern Observatory, Alonso de C\'ordova 3107, Casilla 19001, Santiago 19, Chile
\and
European Southern Observatory, Karl-Schwarzschild-Str. 2, D-85748 Garching, Germany
\and
Astronomical Institute, Slovak Academy of Sciences, 059 60 Tatransk\'a Lomnica, Slovakia
\and
Universit\'e de Nice-Sophia Antipolis, Lab. Lagrange, UMR\,7293, Observatoire de la C\^ote d'Azur, BP 4229, 06304 Nice, France
}
\date{Received 25 September 2012; Accepted 7 February 2013}
\abstract
{The stellar system $\delta$\,Vel contains the brightest eclipsing binary in the southern sky ($\delta$\,Vel~A), and a nearby third star located $\approx 0.6\arcsec$ away ($\delta$\,Vel~B). The proximity of $\delta$\,Vel~B (usable as a reference) makes it a particularly well-suited target to detect the astrometric displacement of the center of light of the eclipsing pair.}
{We obtained NACO astrometric observations with two goals: 1) to confirm the orientation of the orbital plane of the eclipsing pair on the sky determined by interferometry (Paper~III), and 2) to demonstrate the capabilities of narrow-angle adaptive optics astrometry on a simple system with predictable astrometric properties.
}
{We measured the angular separation vector between the eclipsing binary $\delta$\,Vel~A and the visual companion $\delta$\,Vel~B from narrow-band images at 2.17\,$\mu$m obtained with the VLT/NACO adaptive optics system. Based on these observations and our previous determination of the orbital parameters of the wide binary $\delta$\,Vel~A-B, we derived the apparent displacement of the center of light of the eclipsing pair at 11 epochs over its orbital cycle.}
{We detect the astrometric wobble of the center of light of the $\delta$\,Vel~A pair relative to B with a typical measurement precision of $\approx 50\,\mu$as per epoch, for a total amplitude of the measured displacement of $\approx 2$\,mas. }
{The detected wobble is in relatively good agreement with the model we presented in Paper~III, and confirms the orientation of the Aab orbital plane on the sky. The residual dispersion compared to our model is $110\,\mu$as rms, which we tentatively attribute to photometric variability of the fast rotating A-type components Aa and/or Ab in the Br$\gamma$ line. Based on these results, we conclude that in favorable conditions (bright source with only two resolved components, small angular separation), narrow-angle astrometry with adaptive optics on an 8-meter class telescope can reach an accuracy of 50 to 100\,$\mu$as.}
\keywords{Stars: individual: (HD 74956, $\delta$ Vel); Stars: binaries: eclipsing; Methods: observational; Techniques: high angular resolution; Astrometry}

\maketitle

\section{Introduction}

The bright southern star \object{$\delta$ Vel} (\object{HD 74956}, \object{HIP 41913}, \object{GJ 321.3}, \object{GJ 9278}) is a multiple system comprising at least three stars. Its brightest component, $\delta$\,Vel~A, was identified only in 2000 as one of the brightest eclipsing system in the sky \citep{otero00}.
We consider here the two main components of the $\delta$\,Vel system (Aa+Ab and B) as a test case to validate the potential of high-precision narrow-angle astrometry from the ground using adaptive optics. Exciting results were recently obtained using this observing technique, e.g., by \citet{clarkson12}, \citet{hussmann12}, and \citet{koehler12}. These authors typically measured differential astrometry within a field of view of a few arcseconds to an accuracy of a few 100\,$\mu$as \citep[see e.g. Table~4 in][]{clarkson12}. We propose here to push this observing technique to its precision limits on $\delta$\,Vel, which appears to be a particularly favorable target (bright and geometrically simple), in order to evaluate its potential.

In previous works on the $\delta$\,Vel system, \citet[][Paper~I]{kervella09} searched for infrared circumstellar excess in the inner $\delta$\,Vel~A-B system, whose present angular separation is approximately $0.6\arcsec$. Part of the NACO data presented here was also used in this first article, although not for high-precision astrometry.
\citet[][Paper~II]{pribulla11} and \citet[][Paper~III]{merand11} obtained the orbital elements and fundamental parameters (including the parallax of the system) of the three components of the $\delta$\,Vel system.
In the present study, we take advantage of the well-constrained and accurate model of the system obtained in these previous works to predict the astrometric wobble of the center of light (hereafter CL) of the eclipsing system, and compare it with the measured NACO astrometry. We detail our observations and data analysis procedure in Sect.~\ref{observations}, and describe the method we used to derive the wobble of the CL of the eclipsing component in Sect.~\ref{diffastrom}. We discuss in Sect.~\ref{discussion} the differences between the astrometric wobble predicted by the model and the observed displacement.

\section{Observations and data reduction\label{observations}}

\subsection{Observing log and raw data processing}

We observed $\delta$\,Vel at 11 epochs in April-May 2008 and January 2009 using the Nasmyth Adaptive Optics System \citep[NAOS, ][]{rousset03} of the Very Large Telescope (VLT), coupled to the CONICA infrared camera \citep{lenzen98}, which is abbreviated as NACO. These observations were obtained primarily to provide high-precision differential astrometry of the eclipsing system $\delta$\,Vel A relative to B, following the promising demonstration by \citet{seifahrt08}.
We selected the smallest available pixel scale of 13\,mas/pixel, giving a field of view of 13.6$\arcsec$$\times$13.6$\arcsec$. Because of the brightness of $\delta$\,Vel, we employed a narrow-band filter\footnote{http://www.eso.org/sci/facilities/paranal/instruments/naco/} at a wavelength of $2.166 \pm 0.023\,\mu$m (hereafter abbreviated as $2.17\mu$m) together with a neutral density filter (labeled ``{\tt ND2\_short}", that has a transmission of about 1.5\%). The $2.17\mu$m filter wavelength corresponds to the Brackett~$\gamma$ line of hydrogen. A detailed list of the NACO exposures obtained in 2008 is presented in Table~2 of Paper~I. At each epoch 50 exposures were obtained, always centering component A on the same pixel of the CONICA detector array to mitigate the possible effects of field distorsion.

The raw images were processed in a standard way: dark subtraction, flat fielding, and cosmetics removal. We considered three approaches for the flat-field and dark-frame calibration: using specific dark and flat-field frames obtained separately for each night, using average flat/dark frames, and using median flat/dark frames (the last two computed over our 2008 observation epochs). For these different methods, the star positions extracted from the resulting data cubes are identical within $4 \times 10^{-3}$\,pixels, or 50\,$\mu$as. This figure is comparable to our statistical measurement uncertainties, and we conclude that we do not detect a significant astrometric variability induced by the flat field or dark. We therefore chose to use the median flat and dark frames over all observations. This remarkable stability is probably because NACO was kept cold and was not dismounted from the telescope over the period of our 2008 observations. Along this line, \citet{neuhauser08} demonstrated that the plate scale of NACO is extremely stable (to $< 5 \times 10^{-4}$, in relative terms) over three years. The cosmetic corrections we applied concerned only pixels located outside of our astrometric windows (Sect.~\ref{star-separation}), as the areas used for each star position were specifically chosen to present no dead or hot pixels. They therefore had no influence on the astrometric measurements.

\subsection{Star separation measurement \label{star-separation}}

\begin{figure}[]
\centering
\includegraphics[width=7.5cm]{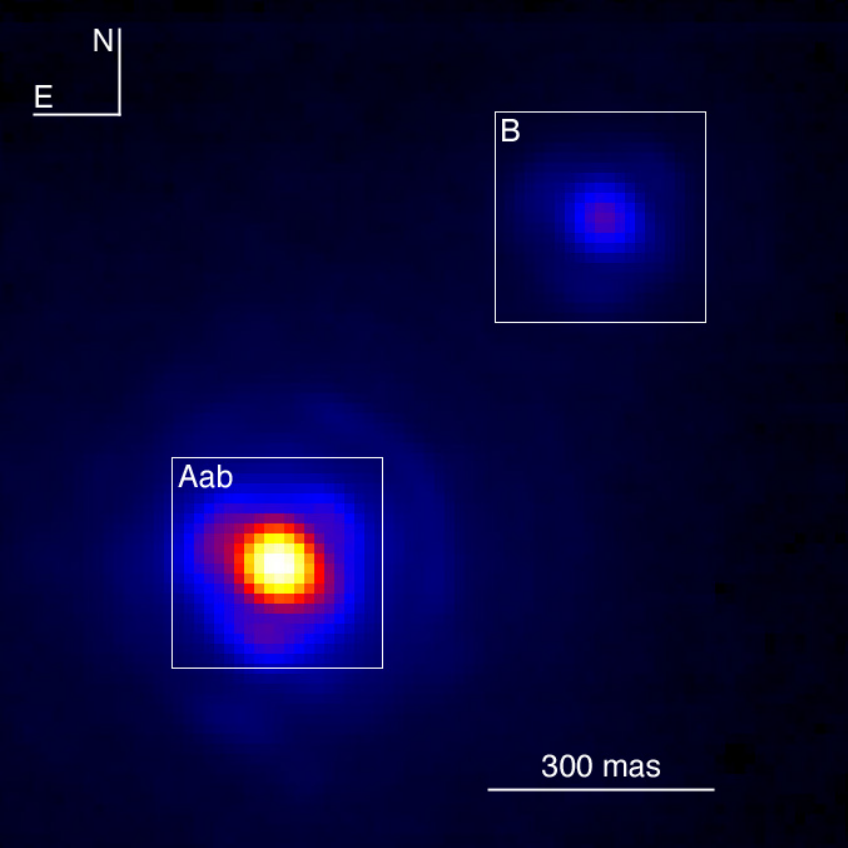}
\caption{Average image of $\delta$\,Vel obtained on 2008 Apr 01, together with the $21 \times 21$\,pix astrometric windows that we used for the differential astrometry (white squares).\label{delvel-boxes}}
\end{figure}

To measure the separation of the two star images on the NACO detector, we treated each image in our data cubes separately. We extracted two sub-images of $21 \times 21$\,pixels centered on the positions of A and B (as shown in Fig.~\ref{delvel-boxes}). We then used a classical $\chi^2$ minimization algorithm to shift and scale the sub-image of $\delta$\,Vel~A over the position of the fainter component B. Our choice of relatively small sub-windows prevents the contamination by the background noise that may present inhomogeneous patterns. We checked that the fit is not sensitive to the chosen size of the sub-windows over a large range of values.
To achieve this fit, we computed a sub-pixel interpolation of the shifted image of A, using the Yorick\footnote{http://yorick.sourceforge.net/} language, and more specifically using an interpolation routine in the Fourier space\footnote{{\tt fft\_fine\_shift} function of the {\tt fft\_utils.i} package}. We derived four parameters: the relative shifts $dx$ and $dy$ in pixels between the two star images, the flux ratio $\rho = f_A/f_B$, and the uniform background level $C$. In order to estimate their associated error bars, we used the bootstrapping technique described in Appendix B of \citet{kervella04}. We validated the Fourier interpolation method by comparing its results with a simple Gaussian fit of the two PSF cores that give the same relative positions (within $150\,\mu$as) although with larger dispersion due to the mismatch of the slightly seeing-distorted PSF and the Gaussian function. Thanks to their angular proximity, $\delta$\,Vel~A and B share the same residual optical aberrations from the adaptive optics system. Because it takes advantage of the similarity of their images, our fitting approach based on Fourier interpolation is particularly efficient.

For the conversion of the star separations measured in pixels to angular values, we adopted the pixel scale of $13.26 \pm 0.03$\,mas/pixel \citep{masciadri03}. This value is in good agreement with the pixel scale obtained by \citet[][$13.24 \pm 0.06$\,mas/pix]{neuhauser08}. Our differential astrometric measurements are affected only by the $\pm 0.23\%$ pixel scale uncertainty over the amplitude of the astrometric wobble of the CL of Aa--Ab and not over the full angular separation of A and B. As shown in Sect.~\ref{diffastrom}, the astrometric wobble amplitude is less than 2\,mas, and the corresponding $\approx 4\,\mu$as systematic error is therefore negligible compared to our statistical accuracy.

\subsection{Differential atmospheric refraction \label{refraction-sect}}

\begin{figure}[]
\centering
\includegraphics[width=7.5cm]{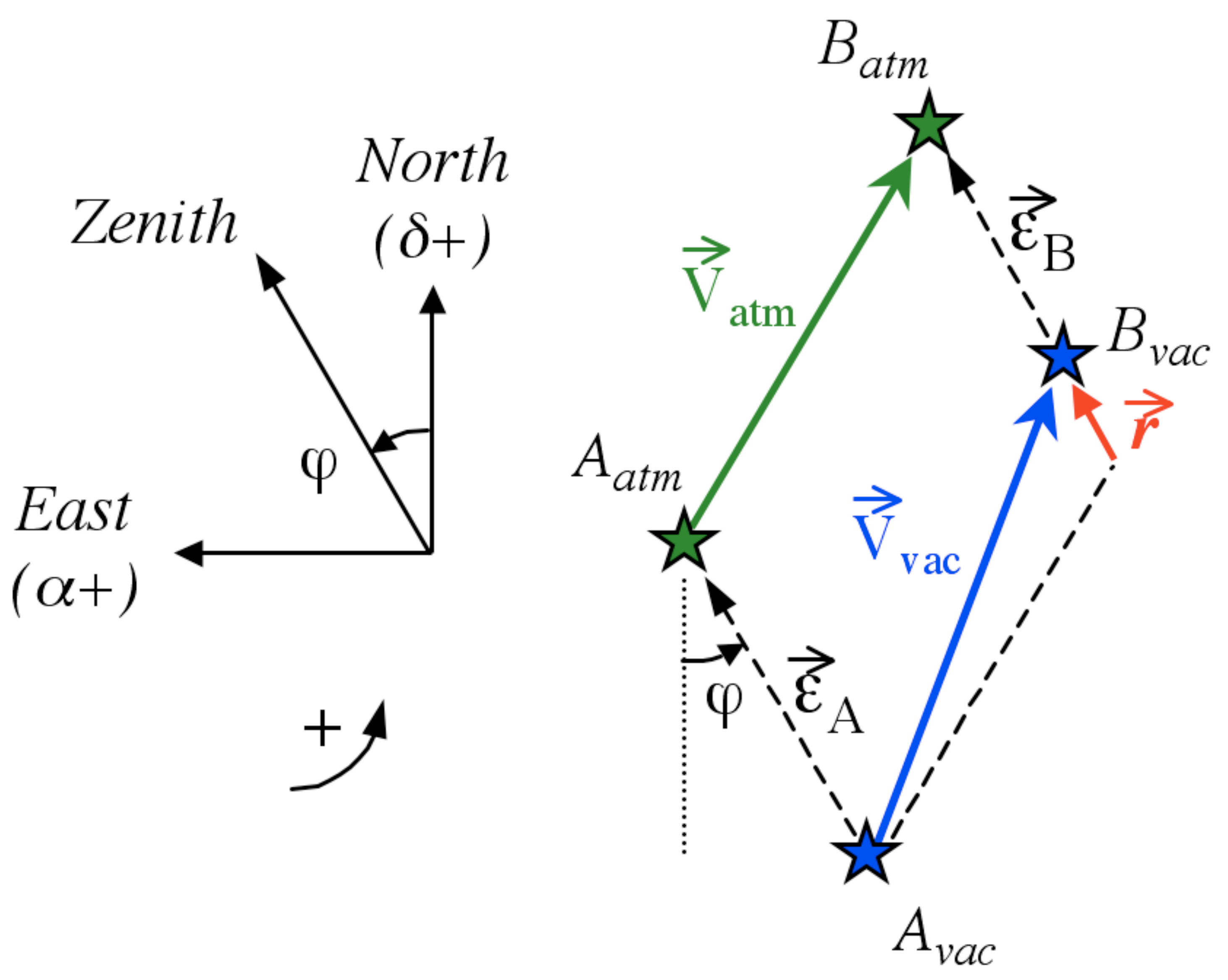}
\caption{Geometrical effect of the atmospheric refraction on the differential astrometry of a binary star (see Sect.~\ref{refraction-sect} for details).
\label{refraction}}
\end{figure}

The angular separation of $\delta$\,Vel~A and B is only $\approx 0.6\arcsec$ for the epoch of our observations. Although this angle is quite small, the effect of differential atmospheric refraction is not negligible considering that the precision of our measurements ($\approx 50\,\mu$as per epoch in right ascension and declination) corresponds to only $\approx 1/12\,000^\mathrm{th}$ of the separation of the two stars.  Figure~\ref{refraction} shows schematically the effect of atmospheric refraction on the apparent positions of the two stars. In this diagram the sky positions of the two stars in the absence of atmosphere are labeled with the {\it ``vac''} subscript, while the apparent positions measured from the ground are labeled {\it ``atm''}; $\varphi$ is the parallactic angle (positive after local celestial meridian crossing) at the time of the observation; $\vec{V_{\rm atm}}$ and $\vec{V_{\rm vac}}$ are the vector separations of A and B with and without atmosphere, respectively; $\vec{\epsilon_A}$ and $\vec{\epsilon_B}$ are the apparent vector displacements of the position of the stars due to atmospheric refraction; and $\vec{r}$ is the resulting differential astrometry correction vector.

The quantity of interest for our purpose is the vector separation of the two stars $\vec{V_{\rm vac}} (\Delta \alpha_{\rm vac}, \Delta \delta_{\rm vac})$. From a simple vectorial closure, the transformation between the measured differential astrometry vector on the sky $\vec{V_{\rm atm}} (\Delta \alpha_{\rm vac}, \Delta \delta_{\rm atm})$ is
\begin{equation}
\vec{V_{\rm vac}} = \vec{V_{\rm atm}} + \vec{r} ,
\end{equation}
where $\vec{r} =  \vec{\epsilon_A} - \vec{\epsilon_B}$.
We computed the values of $|\vec{\epsilon_A}|$ and $|\vec{\epsilon_B}|$ using the {\tt slaRefro} routine from the {\tt Starlink} library\footnote{http://starlink.jach.hawaii.edu/} and the apparent positions of the two stars. The $\vec{r} = (\Delta \alpha_r, \Delta \delta_r)$ vector is always pointing towards zenith. This routine realizes an integration of the effect of the atmospheric crossing on the apparent direction of the incoming star light using a realistic atmospheric model \citep{liebe93}. As input parameters, we used the actual conditions measured at Paranal for each observation epoch (airmass, pressure, temperature, humidity, etc.). Further details on this routine can be found in \citet{gubler98}.
The right ascension and declination corrections to add to the measured NACO differential astrometry of B relative to A are
\begin{equation}
\Delta \alpha_r = |r| \sin \varphi, \ \ \ \ \Delta \delta_r = |r| \cos \varphi .
\end{equation}
For an additional discussion of the effects of atmospheric refraction on adaptive optics observations, the interested reader is referred to \citet{roe02} and \citet{helminiak08}.

\section{Apparent astrometric displacement of $\delta$\,Vel~A\label{diffastrom}}

\begin{table*}
\caption[]{Differential astrometry of $\delta$\,Vel derived from the NACO images. 
\label{diffastrom-table}}
\begin{tabular}{ccccccrrrr}
\hline \hline
UT date & MJD & HJD & $\phi$ & $\theta$ & AM & $\Delta \alpha_r$ & $\Delta \delta_r$ & $[\alpha_B - \alpha_A] \pm \sigma_{\rm stat}$ & $[\delta_B - \delta_A] \pm \sigma_{\rm stat}$ \\
 & -54\,000 & -2\,454\,000 &  & & & & & & \\
\hline
\noalign{\smallskip}
2008 Apr 01 & 557.0224 & 557.5245 & 0.9294 & 0.77 & 1.16 & $-0.014$ & $0.113$ & $-430.570 \pm 0.039 $ & $ 457.819 \pm 0.043$ \\
2008 Apr 04 & 560.9976 & 561.4996 & 0.0175 & 0.96 & 1.16 & $-0.028$ & $0.098$ & $ -430.669 \pm 0.043 $ & $ 457.070 \pm 0.055$ \\
2008 Apr 06 & 562.0121 & 562.5141 & 0.0400 & 0.68 & 1.16 & $-0.010$ & $0.115$ & $-430.828 \pm 0.028 $ & $ 457.046 \pm 0.037$ \\
2008 Apr 07 & 563.0048 & 563.5067 & 0.0619 & 0.92 & 1.16 & $-0.015$ & $0.111$ & $-430.724 \pm 0.028 $ & $ 456.975 \pm 0.038$ \\
2008 Apr 20 & 576.9715 & 577.4732 & 0.3713 & 0.68 & 1.16 & $-0.009$ & $0.113$ & $-430.267 \pm 0.035 $ & $ 452.865 \pm 0.039 $ \\
2008 Apr 23 & 579.0231 & 579.5247 & 0.4167 & 1.02 & 1.19 & $0.038$ & $0.063$ & $-429.964 \pm 0.030 $ & $ 451.974 \pm 0.033$ \\
2008 Apr 24 & 580.9917 & 581.4932 & 0.4603 & 1.30 & 1.16 & $0.026$ & $0.095$ & $-429.790 \pm 0.042 $ & $ 451.090 \pm 0.056$ \\
2008 May 05 & 591.9748 & 592.4761 & 0.7036 & 0.79 & 1.17 & $0.035$ & $0.079$ & $-428.888 \pm 0.046 $ & $ 448.714 \pm 0.042$ \\
2008 May 07 & 593.9732 & 594.4744 & 0.7478 & 0.86 & 1.18 & $0.036$ & $0.073$ & $-429.234 \pm 0.036 $ & $ 447.959 \pm 0.033$ \\
2008 May 18 & 604.0442 & 604.5450 & 0.9709 & 0.76 & 1.48 & $-0.040$ & $-0.013$ & $-429.604 \pm 0.055 $ & $ 445.438 \pm 0.051$ \\
\hline
\noalign{\smallskip}
2009 Jan 07 & 838.1347 & 838.6357 & 0.1556 & 1.20 & 1.41 & $-0.207$ & $0.086$ & $-424.775 \pm 0.084 $ & $ 381.374 \pm 0.116$ \\
\hline
\end{tabular}
\tablefoot{$\phi$ is the orbital phase of the system (relative to periastron passage), $\theta$ the DIMM seeing in arcseconds (in the visible), AM the airmass, and $\Delta \alpha_r$ and $\Delta \delta_r$ the corrections for the differential atmospheric refraction (see Sect.~\ref{refraction-sect} for details). The listed differential astrometry $[\alpha_B - \alpha_A]$ and  $[\delta_B - \delta_A]$ include the atmospheric refraction correction. The angles are all expressed in milliarcseconds.}
\end{table*}

\begin{table}
\caption[]{Photometric flux ratio $f(A)/f(B)$ between A and B in the NACO 2.17\,$\mu$m narrow-band filter. \label{photom-table}}
\begin{tabular}{cccr}
\hline \hline
UT date & MJD & $\phi$ & $f(A)/f(B)$ \\
 & -54\,000 & & \\
\hline
\noalign{\smallskip}
2008 Apr 01 & 557.0224 & 0.9294 & $ 10.547 \pm 0.009 $ \\
2008 Apr 04 & 560.9976 & 0.0175 & $ 10.413 \pm 0.012 $ \\
2008 Apr 06 & 562.0121 & 0.0400 & $ 10.426 \pm 0.008 $ \\
2008 Apr 07 & 563.0048 & 0.0619 & $ 10.332 \pm 0.007 $ \\
2008 Apr 20 & 576.9715 & 0.3713 & $ 10.469 \pm 0.008 $ \\
2008 Apr 23 & 579.0231 & 0.4167 & $ 10.147 \pm 0.012 $ \\
2008 Apr 24 & 580.9917 & 0.4603 & $ 10.589 \pm 0.011 $ \\
2008 May 05 & 591.9748 & 0.7036 & $ 10.295 \pm 0.011 $ \\
2008 May 07 & 593.9732 & 0.7478 & $ 10.269 \pm 0.011 $ \\
2008 May 18 & 604.0442 & 0.9709 & $ 6.993 \pm 0.009 $ \\
2009 Jan 07 & 838.1347 & 0.1556 & $ 10.385 \pm 0.022 $ \\
\hline
\end{tabular}
\tablefoot{The 2008 May 18 epoch corresponds to the primary eclipse.}
\end{table}

The measured relative separations between $\delta$\,Vel~A and B, corrected for differential atmospheric refraction, are listed in Table~\ref{diffastrom-table}. The photometric flux ratio of the two stars is presented in Table~\ref{photom-table}. The orbital phases of the measurements with respect to the eclipsing binary Aab were computed using the following parameters from Paper~II (ROCHE model): $T_0(\mathrm{HJD}) = 2452528.950$ (epoch of periastron passage) and $P=45.15023$\,d. The observed displacement of the CL of Aa--Ab relative to B results from the vectorial sum of the orbital motion of B on its 142-year orbit, and the displacement of the CL of the eclipsing binary system over its $\approx 45$-day orbital cycle. In order to measure this second component only, the secular orbital displacement of $\delta$\,Vel~B relative to A must be subtracted from the measured differential astrometry.

In Paper~III \citet{merand11} determined the orbital elements of the $\delta$\,Vel~A-B system. We used these parameters to estimate the relative positions of the center of mass of $\delta$\,Vel~A and B at each observation epoch, and subtracted the resulting vectors from our NACO astrometric measurements.
The evolution of the measured astrometric position of $\delta$\,Vel~B relative to A as a function of time (over more than a century) is presented in Table~\ref{tableABorbit}. The model positions for the different epochs are also listed, with the residuals compared to the observations.

\onltab{
\begin{table*}
\caption[]{Observed and model positions of $\delta$\,Vel~B relative to A for the epochs presented in Fig.~\ref{delta_Vel_AB_orbit} \citep[see also][]{2002A&A...384..171A}. 
\label{tableABorbit}}
\begin{tabular}{crrrrrr}
\hline \hline
MJD & obs$(\alpha)$ & model$(\alpha)$ & obs$(\delta)$ & model$(\delta)$ & $\Delta \alpha$ & $\Delta \delta$ \\
\hline
\noalign{\smallskip}
13467.7500 &  $0.11 \pm 0.2$ &  0.29308 & $-2.26 \pm 0.2$ & -2.12963 & -0.17874 & -0.12748 \\
15100.3000 &  $0.22 \pm 0.2$ &  0.40581 & $-2.30 \pm 0.2$ & -2.33885 & -0.18842 & 0.03910 \\
18104.7000 &  $0.48 \pm 0.2$ &  0.59401 & $-2.86 \pm 0.2$ & -2.61761 & -0.11537 & -0.24262 \\
19753.3600 &  $0.69 \pm 0.1$ &  0.68520 & $-2.63 \pm 0.1$ & -2.71535 &  0.00502 & 0.08438 \\
21553.8500 &  $0.81 \pm 0.1$ &  0.77407 & $-2.89 \pm 0.1$ & -2.77964 &  0.03269 & -0.10985 \\
24132.7000 &  $0.95 \pm 0.1$ &  0.88042 & $-2.88 \pm 0.1$ & -2.79750 &  0.06595 & -0.08092 \\
25568.0800 &  $0.85 \pm 0.1$ &  0.92825 & $-2.67 \pm 0.1$ & -2.77070 & -0.07696 & 0.10325 \\
27996.4500 &  $1.02 \pm 0.1$ &  0.98936 & $-2.74 \pm 0.1$ & -2.66705 &  0.03325 & -0.06803 \\
31481.5500 &  $0.97 \pm 0.1$ &  1.02966 & $-2.32 \pm 0.1$ & -2.39249 & -0.06104 & 0.07692 \\
34342.1300 &  $0.99 \pm 0.1$ &  1.01591 & $-1.99 \pm 0.1$ & -2.05765 & -0.02202 & 0.06694 \\
34418.8800 &  $0.92 \pm 0.1$ &  1.01491 & $-2.00 \pm 0.1$ & -2.04731 & -0.09296 & 0.05202 \\
34378.0000 &  $0.96 \pm 0.1$ &  1.01545 & $-1.99 \pm 0.1$ & -2.05283 & -0.05359 & 0.06313 \\
43828.7000 &  $0.565 \pm 0.01$ &  0.57324 & $-0.277 \pm 0.01$ & -0.27270 & -0.00838 & -0.00402 \\
43862.7000 &  $0.567 \pm 0.01$ &  0.57025 & $-0.262 \pm 0.01$ & -0.26477 & -0.00289 & 0.00261 \\
48074.5000 &  $0.116 \pm 0.01$ &  0.10948 &  $0.716 \pm 0.01$ &  0.69105 &  0.00643 & 0.02462 \\
48476.0000 &  $0.058 \pm 0.01$ &  0.05763 &  $0.755 \pm 0.01$ &  0.76601 &  0.00045 & -0.01124 \\
48805.0000 &  $0.008 \pm 0.01$ &  0.01475 &  $0.774 \pm 0.01$ &  0.82238 & -0.00664 & -0.04842 \\
51464.4300 & $-0.2886 \pm 0.005$ & -0.30349 &  $1.0200 \pm 0.005$ &  1.00667 &  0.01487 & 0.01328 \\
54581.5100 & $-0.4259 \pm 0.001$ & -0.42983 &  $0.4541 \pm 0.001$ &  0.45143 &  0.00391 & 0.00262 \\
54557.0224 & $-0.4306 \pm 0.001$ & -0.43010 &  $0.4578 \pm 0.001$ &  0.45796 & -0.00053 & -0.00021 \\
54560.9976 & $-0.4307 \pm 0.001$ & -0.43005 &  $0.4571 \pm 0.001$ &  0.45690 & -0.00068 & 0.00020 \\
54562.0121 & $-0.4309 \pm 0.001$ & -0.43004 &  $0.4570 \pm 0.001$ &  0.45663 & -0.00082 & 0.00033 \\
54563.0048 & $-0.4308 \pm 0.001$ & -0.43003 &  $0.4569 \pm 0.001$ &  0.45637 & -0.00073 & 0.00051 \\
54576.9715 & $-0.4303 \pm 0.001$ & -0.42988 &  $0.4528 \pm 0.001$ &  0.45264 & -0.00044 & 0.00013 \\
54579.0231 & $-0.4301 \pm 0.001$ & -0.42986 &  $0.4520 \pm 0.001$ &  0.45209 & -0.00019 & -0.00012 \\
54580.9917 & $-0.4300 \pm 0.001$ & -0.42984 &  $0.4512 \pm 0.001$ &  0.45157 & -0.00013 & -0.00042 \\
54591.9748 & $-0.4290 \pm 0.001$ & -0.42972 &  $0.4485 \pm 0.001$ &  0.44863 &  0.00073 & -0.00010 \\
54593.9732 & $-0.4294 \pm 0.001$ & -0.42969 &  $0.4481 \pm 0.001$ &  0.44809 &  0.00030 & -0.00004 \\
54604.0442 & $-0.4296 \pm 0.001$ & -0.42958 &  $0.4455 \pm 0.001$ &  0.44540 & -0.00005 & 0.00010 \\
54838.1347 & $-0.4246 \pm 0.001$ & -0.42611 &  $0.3813 \pm 0.001$ &  0.38198 &  0.00155 & -0.00070 \\
\hline
\end{tabular}
\tablefoot{All angles are expressed in arcseconds. The model positions were computed using the orbital elements in Paper~III.}
\end{table*}
} 

Because of uncertainties in the determination of the orbital parameters of the A-B system, we added a constant astrometric shift ($\Delta \alpha = -0.3$\,mas, $\Delta \delta = 0$\,mas) to our CL displacement model to match the measured CL positions. To determine this constant shift, we minimized the dispersion of the 2008 measurements with respect to the model. We needed this correction, as our NACO measurements have a limited absolute accuracy caused by the plate scale uncertainty ($\approx 2 \times 10^{-3}$). For the $0.6\arcsec$ separation of $\delta$\,Vel~A and B, this systematic uncertainty is $\approx 1$\,mas and it does not average out with the number of measurements.
We obtained a residual rms dispersion of the NACO measurement points of 110\,$\mu$as compared to our model (excluding the 2008 May 05, 2008 May 07, and 2009 Jan 07 epochs, see Sect.~\ref{discrepant}).

\begin{figure}[]
\centering
\includegraphics[width=\hsize]{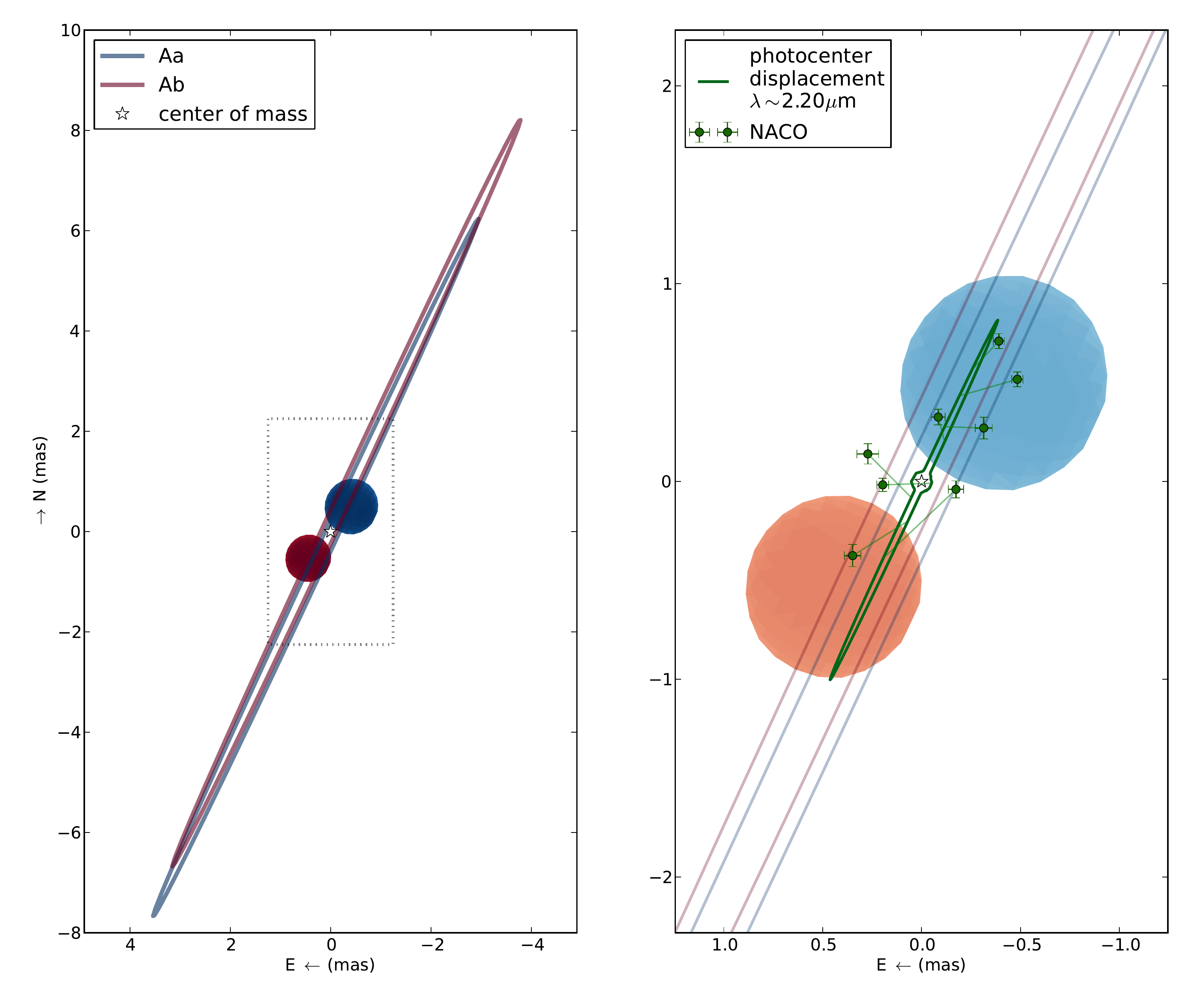}
\caption{{\it Left:} Visual rendering of the Aab system close to the primary eclipse \citep[see also][]{merand11}. {\it Right:} Observed astrometric displacement of the center of light of the Aab system from NACO differential astrometry (points) with the model trajectory of the center of light (thick curve). The orbital trajectories of the Aa and Ab components are represented using thin curves. The rms residual dispersion of the measurements compared to the model is $110\,\mu$as (excluding the 2008 May 05, 2008 May 07 and 2009 Jan 07 epochs, that are also not shown on this plot, see Sect.~\ref{discrepant} for details). \label{delvel-wobble}}
\end{figure}

\begin{figure}[]
\centering
\includegraphics[width=\hsize]{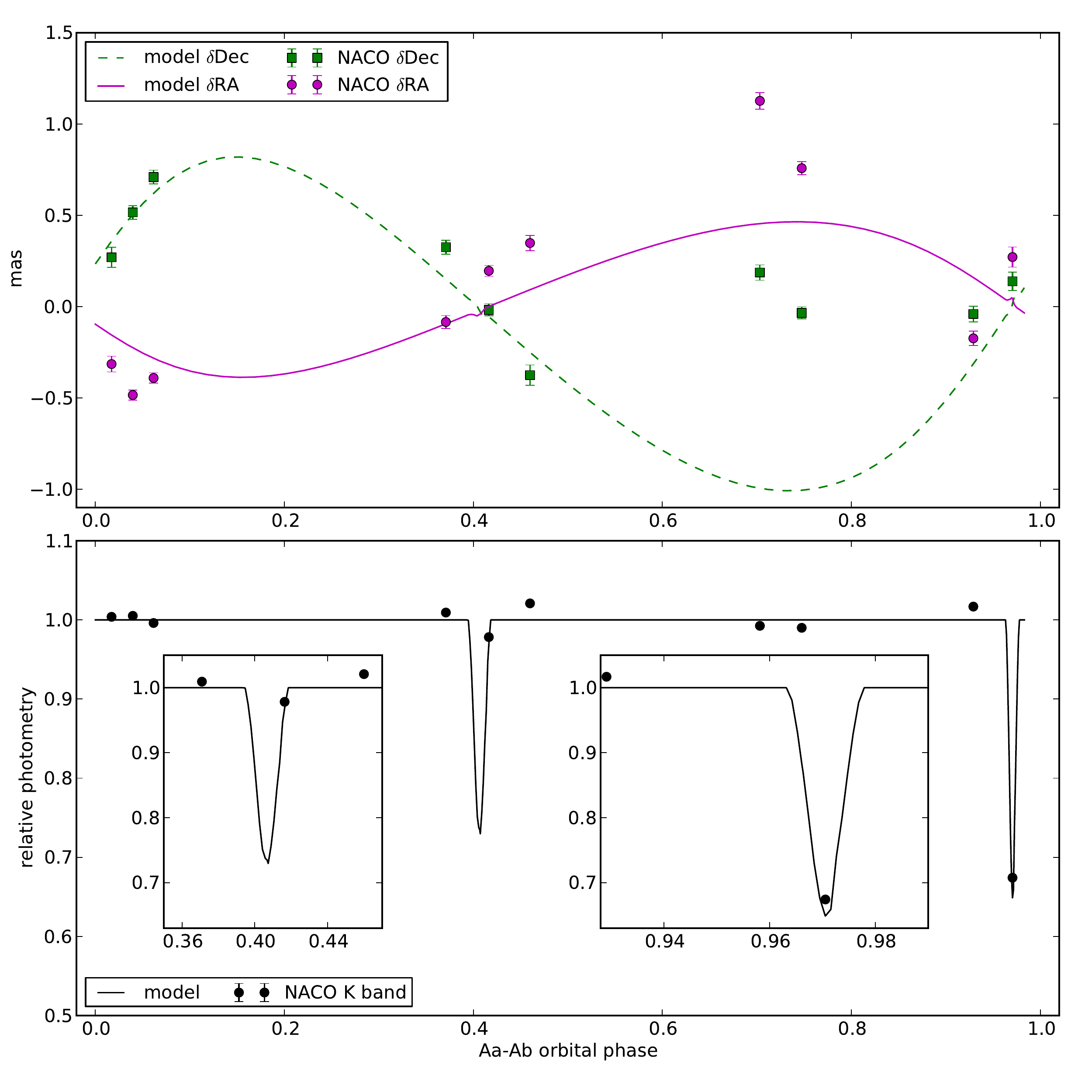}
\caption{{\it Top:} Apparent displacement of the center of light of the $\delta$\,Vel~Aab eclipsing pair relative to B as a function of the orbital phase (the secular displacement of B has been subtracted). The two discrepant measurements around $\phi=0.7$ are discussed in Sect.~\ref{discrepant}.
{\it Bottom}: Ratio of the photometric fluxes of $\delta$\,Vel~A and B in the NACO narrow-band 2.17\,$\mu$m filter (Table~\ref{photom-table}), normalized to the median measured ratio (10.372). For both plots, the solid curves are the predictions of the model presented in Paper~III, and not the result of a fit to the data. \label{delvel-radec}}
\end{figure}

The measured astrometric wobble of the CL of the eclipsing system is presented in Fig.~\ref{delvel-wobble} and the corresponding numerical values are listed in Table~\ref{orbit-model}. Another view of this apparent displacement, including a plot of the differential photometry of $\delta$\,Vel~A-B, is shown in Fig.~\ref{delvel-radec}.
In these two figures, the prediction from the physical model of the eclipsing pair presented in Paper~III is also shown for reference. The derived orientation of the orbital plane of the eclipsing pair is consistent with the interferometric measurements reported in Paper~III. The observed amplitude of the CL displacement is also in satisfactory agreement with the predicted value ($\approx 2$\,mas).

It is interesting to note that at the phases of the eclipses in the model curves, the CL of Aab moves over small arcs because of the slight inclination of the orbital plane on the line of sight and also because of the inhomogeneous light distribution on the apparent disks of the two stars (they are both relatively fast rotators and therefore show some gravity darkening). The amplitude of this effect is too small, however, to be detectable in our NACO measurements.

\begin{table*}
\caption[]{Astrometric displacement of the center-of-light (CL) of $\delta$\,Vel~A. \label{orbit-model}}
\begin{tabular}{ccccccl}
\hline \hline
UT date & MJD & $\phi$ & Model A-B & Model CL & Observ. CL & $N \sigma$ \\
 & - 54\,000 & & vector & displacement & displacement & \\
\hline
\noalign{\smallskip}
2008 Apr 01 & 557.0224 & 0.9294 & $-430.397, 457.959$ & $+0.195, -0.390$ & $-0.173_{\pm 0.039}, -0.140_{\pm 0.043}$ & 10 \\
2008 Apr 04 & 560.9976 & 0.0175 & $-430.355, 456.900$ & $-0.117, +0.278$ & $-0.314_{\pm 0.043}, +0.170_{\pm 0.055}$ & 4 \\
2008 Apr 06 & 562.0121 & 0.0400 & $-430.344, 456.630$ & $-0.191, +0.434$ & $-0.484_{\pm 0.028}, +0.416_{\pm 0.037}$ & 8 \\
2008 Apr 07 & 563.0048 & 0.0619 & $-430.334, 456.365$ & $-0.255, +0.566$ & $-0.390_{\pm 0.028}, +0.610_{\pm 0.038}$ & 4 \\
2008 Apr 20 & 576.9715 & 0.3713 & $-430.183, 452.639$ & $-0.117, +0.201$ & $-0.084_{\pm 0.035}, +0.226_{\pm 0.039}$ & 1 \\
2008 Apr 23 & 579.0231 & 0.4167 & $-430.160, 452.091$ & $-0.051, -0.009$ & $-0.196_{\pm 0.030}, -0.117_{\pm 0.033}$ & 8 \\
2008 Apr 24 & 580.9917 & 0.4603 & $-430.139, 451.565$ & $+0.070, -0.201$ & $+0.349_{\pm 0.042}, -0.475_{\pm 0.056}$ & 7 \\
2008 May 05 & 591.9748 & 0.7036 & $-430.015, 448.627$ & $+0.449, -0.984$ & $+1.127_{\pm 0.046}, +0.087_{\pm 0.042}$ & 28 \\
2008 May 07 & 593.9732 & 0.7478 & $-429.992, 448.093$ & $+0.464, -1.006$ & $+0.758_{\pm 0.036}, -0.134_{\pm 0.033}$ & 26 \\
2008 May 18 & 604.0442 & 0.9709 & $-429.876, 445.399$ & $+0.052, -0.082$ & $-0.272_{\pm 0.055}, +0.039_{\pm 0.051}$ & 5 \\
\hline
\end{tabular}
\tablefoot{``Model A-B vector'' is the predicted astrometric separation $([\alpha_B - \alpha_A], [\delta_B - \delta_A])$ of the centers of light of A and B (with a shift of $\Delta \alpha = -0.3$\,mas, see Sect.~\ref{diffastrom}). The shift of the center of light of Aab relative to its center of mass is listed in the column ``Model CL displacement''. The observed Aab center of light displacement (``Observ. CL displacement'') and the distance between the observed positions and the model positions in number of standard deviations ($N \sigma$) are also listed. All angular values are expressed in milliarcseconds.}
\end{table*}

\begin{figure}[]
\centering
\includegraphics[width=\hsize, angle=0]{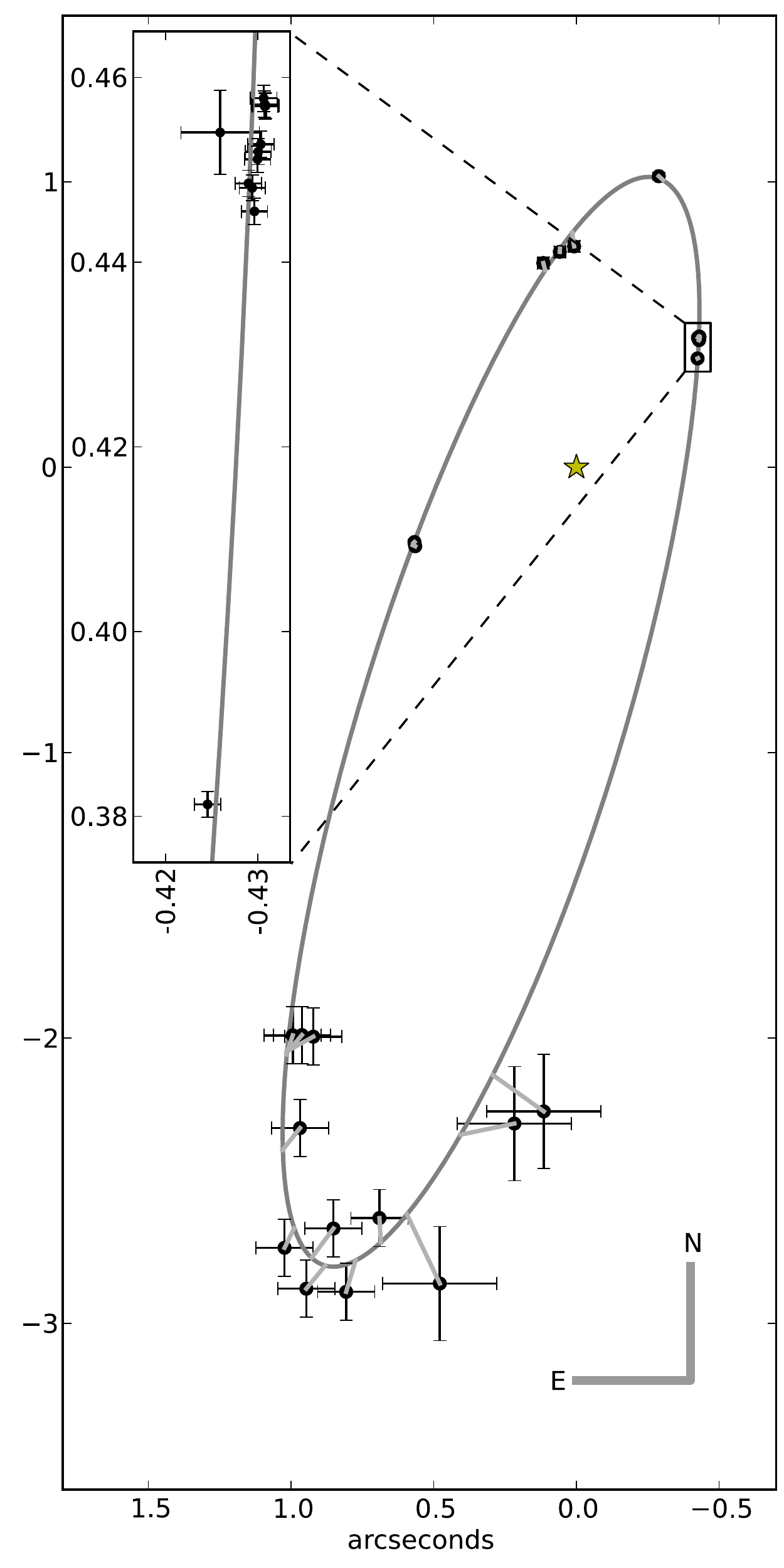}
\caption{Relative orbit of $\delta$\,Vel~B around $\delta$\,Vel~A. The trajectory computed using the orbital elements determined in Paper~III is represented as a black ellipse, and the measurements corresponding to our NACO and VISIR astrometry are shown in the enlarged insert box. The axes are labeled in arcseconds, and the numerical values of the positions are listed in Table~\ref{tableABorbit}. \label{delta_Vel_AB_orbit}}
\end{figure}

%

\section{Discussion\label{discussion}}

\subsection{Inconsistent NACO measurements\label{discrepant}}

The overall agreement between our NACO astrometric measurements and the model of the eclipsing pair is satisfactory.
However, three of our measurement points show a significant departure from the expected CL trajectory, at epochs 2008 May 05, 2008 May 07, and 2009 Jan 07.

For our 2009 epoch, the observed discrepancy is most probably due to the combination of two causes. The first is the instability of the field orientation of NACO over a period of more than 7 months. From the analysis of archival NACO reference astrometric fields, G. Chauvin (private communication) observed that a shift of $0.30 \pm 0.03$\,deg in the true North orientation occured between April and August 2008. We identified an intervention on the instrument in early July 2008 that explains the observed shift (although it does not affect our April-May data). Other actions took place on the instrument later in 2008, that possibly modified or removed part of this rotation shift. The second cause is that the true accuracy of the orbital trajectory of B in our model (Fig.~\ref{delta_Vel_AB_orbit}) is limited by the accumulation of errors with time. Because of the uncertainties on the A-B orbital parameters, it is practically impossible to relate our 2009 measurement unequivocally to the 2008 epochs at a level of a few $10\,\mu$as. We therefore excluded the isolated 2009 epoch from the present astrometric analysis.

For the two epochs in early May 2008, however it is more difficult to reach a solid conclusion on the origin of the deviation.
We considered the following possible explanations:

$\bullet$ {\it Atmospheric or instrumental effect.} We checked the different observational parameters of all the observations, either environmental (seeing conditions, airmass, etc.) or instrumental (flat or bias properties, etc.), but we could not find a significant particularity of the data collected over these two nights. The observed deviation was present over two nights separated by only two days, and was not present before or after these two nights. Although a peculiar instrumental bias cannot be formally excluded, this points to an astrophysical origin.

\begin{figure}[]
\centering
\includegraphics[width=\hsize]{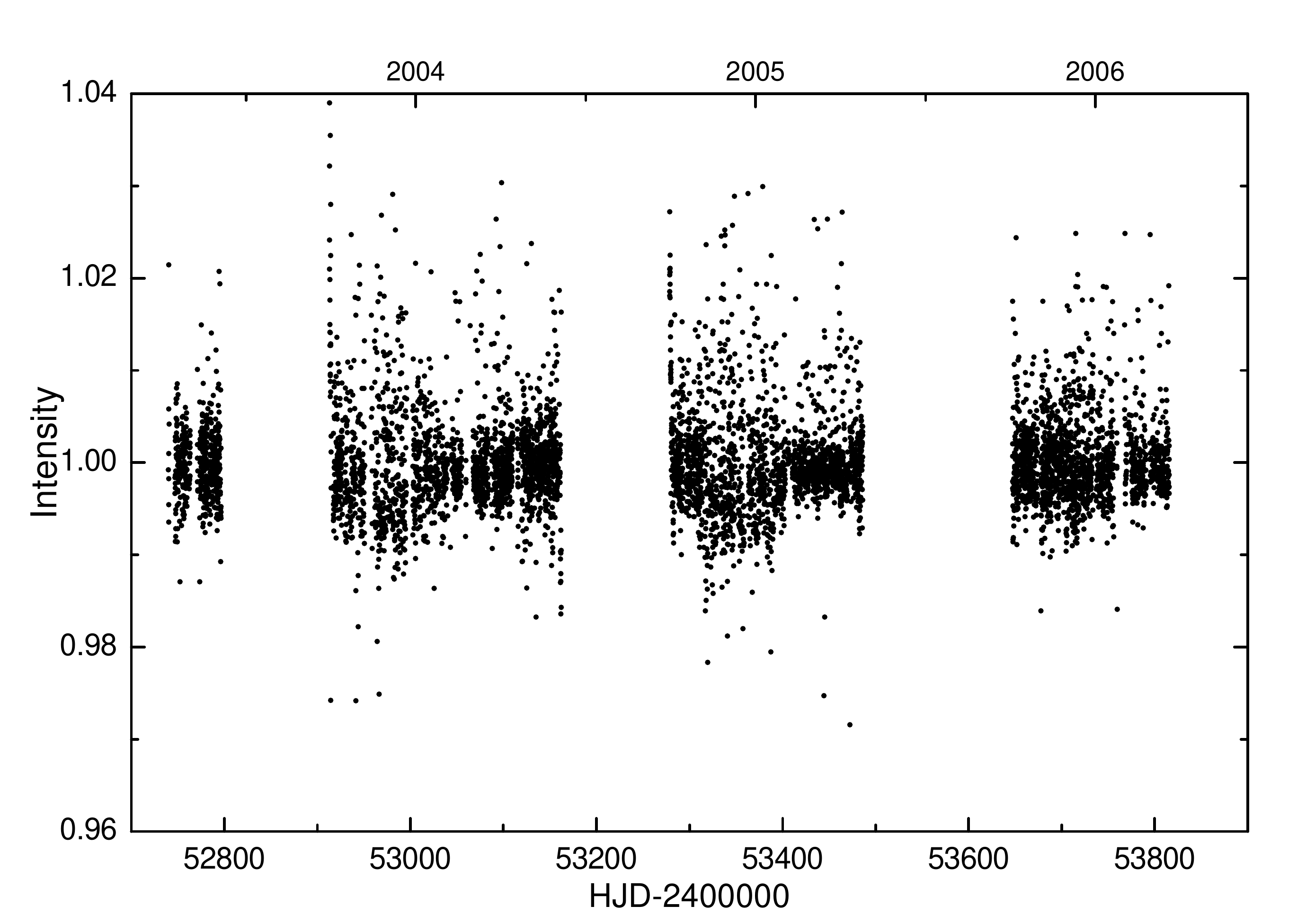}
\caption{Total flux of the $\delta$\,Vel~A-B system as a function of the phase from SMEI observations, excluding the eclipses, normalized to its average over the observation period.\label{delvel-smei}}
\end{figure}

$\bullet$ {\it Transit of a dark body (exoplanet or faint star).} The geometry of the $\delta$\,Vel~A-B eclipsing system makes it a favorable configuration for the transit of dark bodies in front of the stars of the eclipsing pair.
The existence of these systems was predicted by \citet{schneider94}, and recently demonstrated by \citet{armstrong12}.
A search for transit signatures in the photometry measured with SMEI (Fig.~\ref{delvel-smei}) did not give positive results down to about 2\% of the total Aa+Ab+B flux. There is still a possibility that a very shallow transit of 1-2\% of the total flux may be present ($\approx 4-5$\% of the flux of one of the main components Aa or Ab), but it would be insufficient to explain the observed astrometric signal. So the transit of a dark component in front of one of the stars appears unlikely.

$\bullet$ {\it Activity of $\delta$\,Vel Aa and/or Ab in the Br$\gamma$ line.} The NACO images were taken with a narrow-band Br$\gamma$ filter.
It is therefore possible that episodic mass loss created a temporary envelope and circumstellar emission around one of the two eclipsing stars. An emission of this kind could be faint in the broad-band photometry of the system (SMEI data), but would be much more apparent in our narrow-band Br$\gamma$ filter. This would affect our astrometric measurements by displacing the CL of the eclipsing system towards the star experiencing the circumstellar emission.
According to our model of the system (Paper~III), the expected Br$\gamma$ flux ratio between $\delta$\,Vel~A and B is 10.67. We measure a median value of $f(A)/f(B) = 10.37 \pm 0.13$ over all our observations (Table~\ref{photom-table}), which is slightly below the model, but could be explained by interstellar extinction. It should be noted that the flux ratio values observed on 2008-05-05 and 05-07 are both lower than the median value (10.27 to 10.30). Moreover, the observed dispersion of the measurement points ($\approx 0.13$) is much larger than the uncertainty of the measurements ($\approx 0.01$). This means that either Aa, Ab, or B experiences photometric variability in the Br$\gamma$ line. The SMEI photometry presented in Fig.~\ref{delvel-smei} also shows a possible variability at a 2-3\% level. If we consider only the case of a change in brightness of one of the eclipsing stars (a change in the flux of B would have no effect on the astrometry), a 1\% change in total flux means a $\approx 2$\% change in the relative brightness of $\delta$\,Vel~Aa and Ab. This change in flux would result in a geometrical shift of the CL of the system of approximately $300\,\mu$as when the stars are at maximum elongation (separation $\approx 16$\,mas). This amplitude is similar to the observed drift of the CL on 2008 May 05 and 2008 May 07. However, the fact that we have only the relative photometry of A and B at high precision means that we cannot exclude a change in the brightness of B that would invalidate this hypothesis.

$\bullet$ {\it Binarity of $\delta$\,Vel~B.} The presence of an undetected companion around B would cause a displacement of the photocenter of this component, and thus affect the differential A-B astrometry. The mass of the companion of B would have to be relatively high, and its orbital period short, but we cannot formally exclude this possibility. As suggested in Paper~II, spatially resolved spectroscopy of $\delta$\,Vel~B would be necessary to check this hypothesis, as well as to confirm the physical properties of this star.

In summary, the most likely explanations for the inconsistent astrometric measurements of 2008 May 05 and 2008 May 07 are either an unidentified instrumental bias, or a transient activity of one of the fast rotating Aa or Ab stars in the Br$\gamma$ hydrogen line (e.g. the temporary formation or disappearance of a disk contributing $\approx 1-2$\% of the flux in the Br$\gamma$ line). The sensitivity of our differential astrometric measurements to the flux ratio of Aa and Ab introduces an additive noise on the measurement of the Aab CL displacement. Although the formal accuracy of each individual astrometric measurement is unaffected, the level of agreement with the CL displacement predicted by our model (that assumes non-variable stars) is reduced.

\subsection{Astrometric accuracy}

The dispersion of our NACO astrometric measurements around our model of the system ($110\,\mu$as rms, excluding the 2008 May 05 and 2008 May 07 epochs, see Fig.~\ref{delvel-wobble}) is larger than our individual measurement uncertainty ($\approx 50\,\mu$as per epoch). We therefore estimate that the true accuracy of a single measurement epoch is most probably between 50 and 100\,$\mu$as. This level of accuracy is consistent with the conclusions of \cite{fritz10}.
This confirms that very high-precision astrometric observations are feasible from the ground with adaptive optics systems on large telescopes. The measurements we presented were obtained on a particularly favorable object, very bright and with only two unresolved objects in the field of view. However, the detected astrometric displacement is particularly subtle, and the fact that we could retrieve the orientation of the orbital plane of the eclipsing pair demonstrates the potential of this technique.

One should also note that we used a neutral density filter and a narrow band filter to prevent the saturation of the CONICA detector. This means that the level of accuracy we demonstrated is directly applicable to significantly fainter targets than $\delta$\,Vel (probably by at least 10 magnitudes). For even fainter objects ($m_K > 14$), particularly in the Galactic Center region, a detailed discussion is presented by \citet{fritz10}.

\section{Conclusion}

We reported high precision differential astrometric measurements of the triple system $\delta$\,Vel. They show that it is possible to measure the separation of close binaries of suitable brightness with a repeatable precision of 50 to 100 $\mu$as. The comparison of the measured astrometric displacement of the CL of the eclipsing component of $\delta$\,Vel with a well-constrained model of the system shows that our NACO measurements give a realistic astrometric signal, in terms of amplitude and orientation on the sky. However, this qualitatively good agreement is affected by what we interpret as photometric variability of one of the components of the eclipsing binary (although other explanations are also possible).

High accuracy astrometric measurements have the potential to bring novel constraints in several fields of astrophysics, from the Galactic center \citep[e.g.][]{gillessen09}, to faint companion detection \citep[e.g.][]{tokovinin12}, or cluster dynamics \citep[e.g.][]{clarkson12, hussmann12}.
The future instrumentation of extremely large telescopes \citep[see e.g.][]{trippe10} will give access to much fainter objects, although the actual measurement accuracy will probably not be improved significantly beyond the 50\,$\mu$as limit.
The next generation interferometric instrumentation at near-infrared wavelengths, particularly the GRAVITY \citep{gillessen10,eisenhauer11,vincent11} and ASTRA \citep{woillez12,stone12} instruments, are expected to overcome this limit, down to an accuracy of 10 to $20\,\mu$as over a field of view of a few arcseconds.

\begin{acknowledgements}
We thank the La Silla Paranal Observatory team for the successful execution of the NACO observations of $\delta$\,Vel.
The visual binary orbit fitting procedures we used were created by Dr Pascal Bord\'e.
We thank Drs Ga{\"e}l Chauvin and Julien Girard for their help in characterizing the stability of the astrometric calibration of NACO.
This work received the support of PHASE, the high angular resolution partnership between
ONERA, Observatoire de Paris, CNRS and University Denis Diderot Paris 7.
This work was supported by the Slovak Research and Development Agency under the contract No. APVV-0158-11, and partially supported by the VEGA 2/0094/11 project.
This research took advantage of the SIMBAD and VIZIER databases at the CDS, Strasbourg (France), and NASA's Astrophysics Data System Bibliographic Services.
\end{acknowledgements}

{}

\end{document}